\documentclass[article,aps,prb, twocolumn,showpacs,amsmath,amssymb,superscriptaddress,nofootinbib,floatfix]{revtex4-1}

\usepackage{color}         
\usepackage[pdftex]{graphicx}      
\usepackage{bm}
\PassOptionsToPackage{square,comma,numbers,sort&compress,super}{natbib}
\usepackage{natbib}
\setcitestyle{square, comma, numbers,sort&compress, super}
\usepackage[colorlinks,plainpages=false,linkcolor=black,urlcolor=black,citecolor=black,pdfpagemode=UseNone,pdfstartview=FitBH]{hyperref}
\usepackage{float}
\usepackage{placeins}
\usepackage{achemso}
\usepackage{tikz}
\usepackage{color}
\usetikzlibrary{shapes}
\definecolor{mygreen}{rgb}{0,0.5,0}
\definecolor{mybrown}{RGB}{165,42,42}
\definecolor{mymagenta}{RGB}{255,0,255}
\definecolor{myyellow}{RGB}{255,255,0}
\definecolor{mycyan}{RGB}{0,255,255}

\begin{document}

\title{Charge Ordering in Superconducting Copper Oxides}

\author{Alex Frano}
\email{afrano@ucsd.edu}
\affiliation{Department of Physics, University of California, San Diego, California 92093, USA}

\author{Santiago Blanco-Canosa}
\email{sblanco@dipc.org}
\affiliation{Donostia International Physics Center, DIPC, 20018 Donostia-San Sebastian, Basque Country, Spain}
\affiliation{IKERBASQUE, Basque Foundation for Science, 48013 Bilbao, Basque Country, Spain}

\author{Bernhard Keimer}
\email{b.keimer@fkf.mpg.de}
\affiliation{Max Planck Institute for Solid State Research, Heisenbergstr. 1, 70569 Stuttgart, Germany}

\author{Robert J. Birgeneau}
\email[]{robertjb@berkeley.edu}
\affiliation{Department of Physics, University of California, Berkeley, California 94720, USA}
\affiliation{Department of Materials Science and Engineering, University of California Berkeley, Berkeley, CA 94720, USA.}


\begin{abstract}
Charge order has recently been identified as a leading competitor of high-temperature superconductivity in moderately doped cuprates. We provide a survey of universal and materials-specific aspects of this phenomenon, with emphasis on results obtained by scattering methods. In particular, we discuss the structure, periodicity, and stability range of the charge-ordered state, its response to various external perturbations, the influence of disorder,  the coexistence and competition with superconductivity, as well as collective charge dynamics. In the context of this journal issue which honors Roger Cowley's legacy, we also discuss the connection of charge ordering with lattice vibrations and the central-peak phenomenon. We end the review with an outlook on research opportunities offered by new synthesis methods and experimental platforms, including cuprate thin films and superlattices.
\end{abstract}

\maketitle

\subsection{Introduction.}

\begin{figure}
  \includegraphics[width=0.5\textwidth]%
    {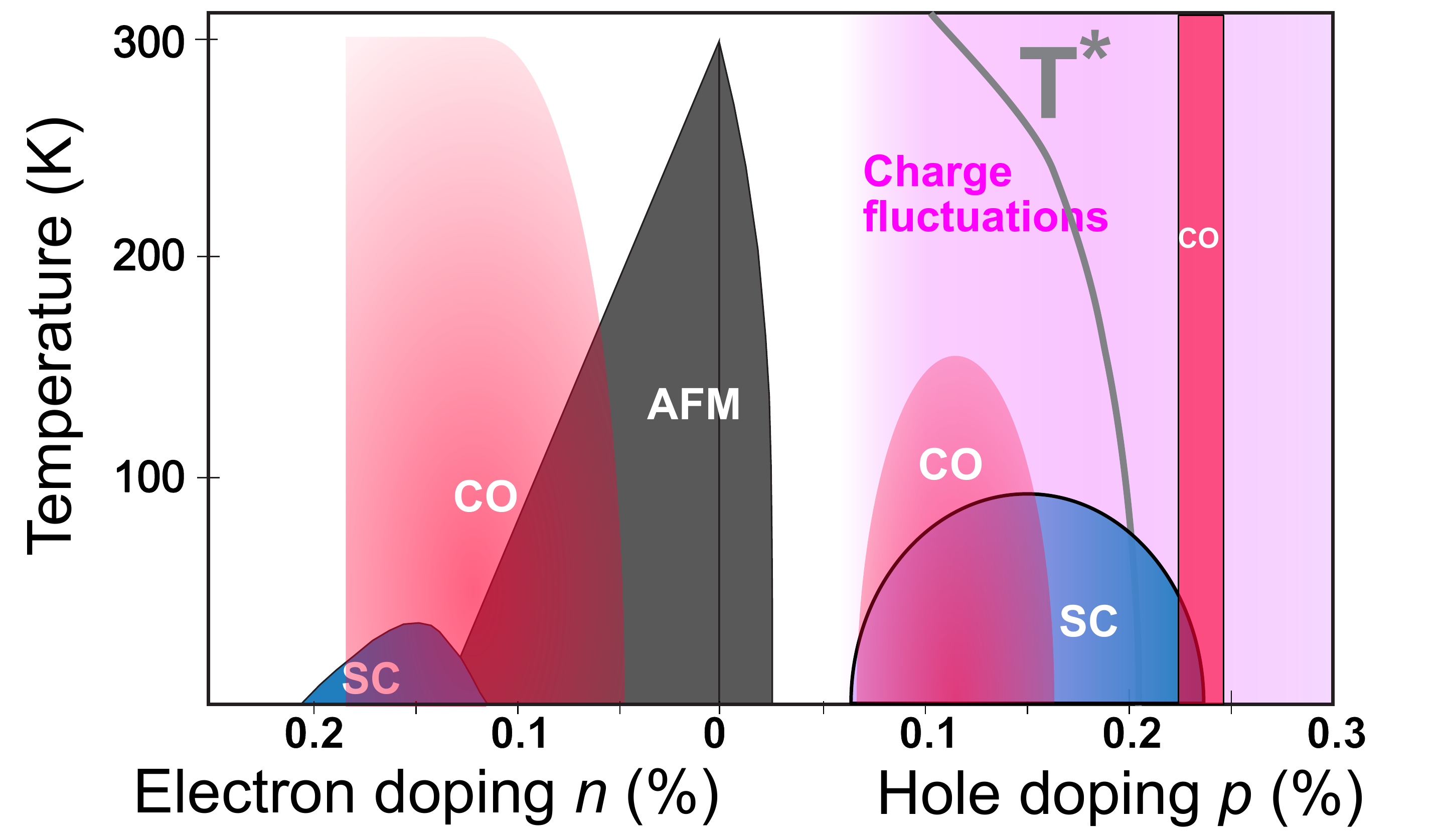}
     \caption{\textbf{A contextual summary of CO physics in cuprates.} The phase diagram of doped cuprates, showing antiferromagnetic (AFM), superconducting (SC), charge-ordered (CO) regions. T$^*$ denotes the onset of the pseudogap.}\label{fig:phase_diagram}
\end{figure}

Quantum materials with strongly correlated electrons are a major platform for the investigation of emergent phenomena, which occur when a system's macroscopic properties are governed by the collective behavior of a particle ensemble rather than by its individual components \cite{Keimer2017}. The intricate interplay between static and dynamic spin, charge, orbital, and pair correlations of the valence electrons and their influence on the macroscopic phase behavior of quantum materials continues to intrigue a large community of researchers. Copper oxide superconductors are an archetypical platform for research on emergent quantum phenomena ~\cite{Keimer2015, noman_pseudogap_review,taifeller_cuprate_review}. Despite their simple lattice architecture and band structure (with CuO$_2$ square planes as the only generic electronically active building block, and a single Cu $d$-orbital and two O $p$-orbitals dominating the electronic states near the Fermi level), they exhibit a set of electronic phases with radically different macroscopic properties as a function of the concentration of mobile charge carriers in the CuO$_2$ planes. Among the multiple phases that have been proposed and studied in the cuprates, an antiferromagnetic insulator as well as unconventional metallic and superconducting phases (Figure~\ref{fig:phase_diagram}) have long been accepted as ``universal'' -- that is, they are found in all chemically distinct families of cuprates that can be prepared with the corresponding doping level.

This brief review addresses recent developments in research on charge order, which was recently recognized as another universal feature of the copper oxide phase diagram.
The concept of charge order has a long history in solid-state research. Already in the 1950s, Peierls recognized that the total energy of a one-dimensional metal could be lowered through a lattice distortion with a wave vector matching the Fermi wave vector \cite{Peierls}. Evidence for Peierls distortions (also known as ``charge density waves'', CDWs) was later found in many binary and ternary compounds with weakly correlated quasi-one-dimensional (-1D) or quasi-2D electron systems~\cite{Monceau,TMDC_CDW,Chromium_CDW,Holmium_CDW}. In most quasi-2D systems, CDW formation reconstructs the Fermi surface, but does not remove it entirely. Another paradigm of charge ordering emerged in research on transition metal oxides comprising metal ions with different stable valence states, such as manganates, nickelates, and cobaltates  ~\cite{Staub_CO_nickelates2002,Yi_Lu_CO_nickelates2016,Mori_stripes_manganites,Tranquada19982150,babkevich_Cobaltites_2016}. Depending on the doping level, various superstructures of these valence states have been observed. They tend to be commensurate with the underlying lattice, and they are often associated with commensurate magnetic order~\cite{Khomskii}.

Originally predicted in 1989~\cite{Zaanen1989,MACHIDA1989192}, the experimental discovery of charge order in the cuprates dates back to 1995, when Tranquada and coworkers performed neutron diffraction experiments on a La-based cuprate with a doping level of 1/8 holes per square-lattice unit cell~\cite{Tranquada_stripes_1995}. They found Bragg reflections characteristic of a magnetic superstructure with a periodicity that is eight times larger than the basic crystallographic unit cell, as well as another set of reflections indicating a four-unit-cell modulation of the crystal lattice (presumably in response to a modulation of the valence electron density). Because of the commensurate nature of the charge modulation and the coexistence with magnetic order, the ordered state was discussed in a real-space picture, which subsequently evolved into a new paradigm for charge order -- often referred to as the ``striped phase''. In this picture, nanoscale regions with weakly perturbed antiferromagnetic order are separated by nonmagnetic ``rivers of charge'' where the mobile holes can delocalize without breaking magnetic bonds.

It was soon realized, however, that the striped phase only occurs in La-based superconductors, where it is stabilized by soft tilt distortions of the oxygen octahedra surrounding the Cu ions in this lattice structure~\cite{abbamonte_spatially_2005,Fink_LESCO_1,Fink_PRB83, Hucker_LBCO_PRB_2011,Hucker_LBCO_PRB_2011,Wilkins2011,wu_charge_2012,Jacobsen2015,Fabbris2013}. Neutron diffraction experiments on other cuprates, including YBa$_2$Cu$_3$O$_{6+x}$, did not detect magnetic superstructures associated with the striped phase. However, the similarities of the magnetic excitations in moderately doped cuprates with \cite{Tranquada2004} and without \cite{Bourges2000,Hayden2004} stripe order revealed by inelastic neutron scattering experiments (with important contributions from Roger Cowley and his coworkers \cite{Cowley_2005,Cowley_2010}) stimulated theories of dynamical stripes, where quantum fluctuations obliterate static magnetic order \cite{Kivelson2003}. Another class of theories proposed bond-centered charge order, where the charge modulation predominantly affects the oxygen ions, and spins-1/2 on a Cu-O-Cu bonds form non-magnetic singlets that are not directly visible in neutron diffraction experiments \cite{Vojta_stripes}. This scenario was supported by scanning tunneling spectroscopy (STS) studies of Ca$_{2-x}$Na$_x$CuO$_2$Cl$_2$ and Bi-based superconductors, which found direct evidence of bond-centered charge modulations~\cite{hoffman_four_2002,hanaguri_checkerboard_2004,kohsaka_intrinsic_2007,Wise2008,parker_fluctuating_2010}.These materials, however, are known for their comparatively high level of chemical disorder and inhomogeneity. The charge ordering patterns found by STS are short-range ordered with correlation lengths not exceeding several unit cells, raising the question to what extent they might actually be caused by disorder, in analogy to ``Friedel oscillations'' in elemental metals \cite{PhysRevB.93.205117}. Recent numerical simulations of the doped Hubbard ~\cite{Huang1161,Zheng1155} and $t-J$~\cite{choubey_incommensurate_2017} models have yielded firm evidence of stripe order (both static and fluctuating), with details that depend sensitively on the model parameters and the doping level. The most recent calculations of the Hubbard and $t-J$ models capture some of the interplay between density wave orders and superconductivity~\cite{jiang_superconductivity_2019,jiang_superconductivity_2018}.

Recently, the application of resonant x-ray scattering (RXS) to the cuprates opened up new perspectives on this complex issue. Taking advantage of photons tuned to the dipole-active $L$-absorption edge of Cu, resonant scattering can discriminate between diffraction features originating from modulations of the valence electron density in the CuO$_2$ planes, and diffuse scattering generated by random lattice displacements due to chemical substitution, which dominates non-resonant x-ray scattering maps of most doped cuprates (where both valence and core electrons contribute). RXS experiments found superstructure reflections indicative of charge order in moderately doped YBa$_2$Cu$_3$O$_{6+x}$~\cite{Ghiringhelli2012,Achkar2012}, which is regarded as one of the cuprate families least affected by chemical disorder because the oxygen dopant atoms are arranged in chains of several tens of nanometers length~\cite{Bobroff2002}. Around the same time, nuclear magnetic resonance and non-resonant x-ray diffraction measurements also found evidence of charge order in YBa$_2$Cu$_3$O$_{6+x}$~\cite{chang_2012,Wu2011,Wu2013,Wu2015}. The x-ray data further showed that the wavevector characterizing charge ordering in YBa$_2$Cu$_3$O$_{6+x}$ is incommensurate with the underlying crystal lattice, in qualitative analogy to charge density waves in weakly correlated metals. This observation stimulated a large number of weak-coupling theories on the origin of this phenomenon and its relationship to the Fermi surface inferred from transport measurements on YBa$_2$Cu$_3$O$_{6+x}$ \cite{Met2010,Chubukov2014}. In particular, the presence of electron pockets revealed by Hall effect and quantum oscillation experiments on moderately hole-doped cuprates was attributed to a CDW-induced Fermi surface reconstruction~\cite{Doi07,laliberte_2011,Tabis2014,doiron-leyraud_evidence_2015} and there has been much debate about possible contributions of charge order and charge fluctuations to the spontaneous lattice-rotational symmetry breaking  (``electronic nematicity'') that is widely observed in the underdoped regime~\cite{Fernandes2019}.

Following up on the early work on YBa$_2$Cu$_3$O$_{6+x}$, a large body of experiments has confirmed that charge order is a universal feature of moderately doped cuprates, with a set of properties that are shared between all chemically distinct families of cuprates where this phenomenon has been studied (Table~\ref{tab:corr_lengths}). At the same time, pronounced materials-specific variations in this phenomenology were also found, which is not unexpected because modulations of the valence-electron density
couple to the materials-specific lattice structure via long-ranged Coulomb interactions. Indeed, crystallographic refinements in the charge-ordered state in YBa$_2$Cu$_3$O$_{6+x}$ have shown substantial displacements of nearly all ions in the unit cell \cite{Forgan2015}, and the correlation lengths of charge ordered states in different cuprates suggest a high sensitivity to chemical disorder associated with doping. The relevance of crystallographic disorder was further highlighted by theoretical work demonstrating that 2D incommensurate electronic superstructures are highly sensitive to phason disorder \cite{nie_vestigial_2017}.

In this brief review -- which expands on previous reviews of the subject~\cite{Tranquada_review, Comin_review} -- we will summarize the universal properties of charge order in cuprates including its geometry, temperature dependence, dimensionality and spatial range. We will also discuss materials-specific variations, highlighting recent discoveries whose universality is still under investigation. Honoring Roger Cowley and his legacy of scattering techniques, we will focus on x-ray scattering results and draw parallels to his seminal work on soft-mode structural phase transitions and associated ``central peaks'' ~\cite{Cowley1996_STO,Cowley_2006}.


\newcommand{\markerthree}{\raisebox{0.5pt}{\tikz{\node[draw,scale=0.4,circle,fill=black!20!myyellow](){};}}}
\newcommand{\markerfour}{\raisebox{0.5pt}{\tikz{\node[draw,scale=0.3,regular polygon, regular polygon sides=3,fill=mycyan!0!mycyan,rotate=0](){};}}}
\newcommand{\markerfive}{\raisebox{0.5pt}{\tikz{\node[draw,scale=0.4,regular polygon, regular polygon sides=4,fill=mymagenta](){};}}}

\begin{figure*}
  \includegraphics[width=0.7\textwidth]%
    {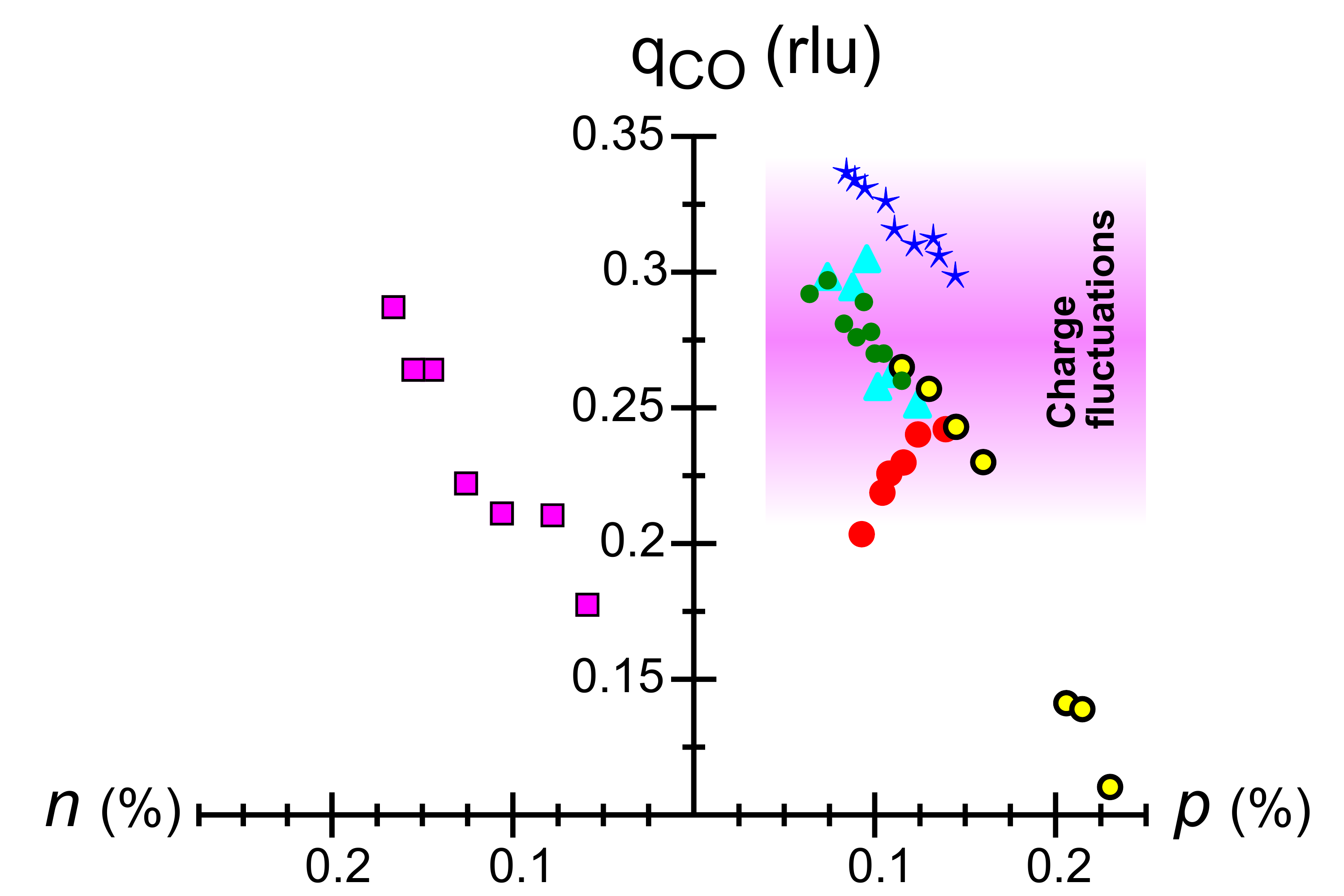}
     
     \caption{\textbf{The momentum versus doping behavior.} The dependence of the charge ordering wavevector $q_{CO}$ as function of doping for various families, denoted in the compounds' respective reciprocal lattice units (rlu) defined along the Cu-O bond direction: YBa$_2$Cu$_3$O$_{6+x}$ (\textcolor{blue}{$\star$}),214-LCO (\textcolor{red}{\textbullet}), Bi-2201 (\protect\markerthree),Bi-2212 (\protect\markerfour), Hg-1201 (\textcolor{mygreen}{\textbullet}), and electron-doped materials (\protect\markerfive). The region in pink denotes the $q$-range where fluctuations are observed in hole-doped materials.  }\label{fig:q_vector}
\end{figure*}

\subsection{Structure and periodicity.}

Over the past few years, all major cuprate families have been investigated by RXS, and evidence of charge order with ordering vector along the Cu-O-Cu bond direction has been found at moderate doping levels in almost all systems (Figure~\ref{fig:phase_diagram} and Table~\ref{tab:CDW_summary}). The sole exception is a glassy, rotationally symmetric state that has recently been observed in lightly electron-doped cuprates \cite{Kang19}. The bond-oriented charge order disagrees with early theories of Fermi surface instabilities in spin-fermion theories~\cite{Met2010}, but later work showed that instabilities matching the experimental observations can also occur in one-band~\cite{Chubukov2014} and three-band~\cite{Thomson2015} models. Alternatively, the charge-ordering wavevector has also been described in a real-space picture in the spirit of the original ``stripe'' model, where it characterizes the periodicity of a nanoscale array of phase-separated regions~\cite{capati_electronic_2015}.

\begin{table*}[htp]
	\begin{tabular}{ || p{3cm} |  p{3.5cm} | p{3.5cm} | p{3cm} | p{2cm}  ||  }
		\hline \hline
		\hspace{0.05cm}\textbf{Cuprate System} & \hspace{0.05cm} \textbf{Low-T $q$-vector range (rlu)$^{\dagger}$} & \hspace{0.05cm} \textbf{High-T fluctuations $q$-vector range (rlu)}   &   \textbf{Dimensionality} & \textbf{CO in overdoped?} \hspace{0.15cm}  \\

		\hline \hline
		\hspace{0.05cm} La$_2$CuO$_{4}$ (214-LCO) & \hspace{0.05cm} 0.2 -- 0.245 \cite{Fink_LESCO_1,Fink_PRB83, Hucker_LBCO_PRB_2011}  & \hspace{0.05cm} $\sim$0.3 \cite{miao_RIXS_2017,miao_IXS_2018} & 2D$^{\diamond}$ \cite{Fink_LESCO_1,Fink_PRB83, Hucker_LBCO_PRB_2011} & Unknown  \\
		\hline
		
		\hspace{0.05cm} YBa$_2$Cu$_3$O$_{6+y}$  & \hspace{0.05cm} 0.33 -- 0.29 \cite{Ghiringhelli2012,chang_2012,Blanco2013,Blackburn2013,Bluschke2019}  & \hspace{0.05cm}  $\in(\sim0.2,\sim0.4)$ \cite{arpaia_dynamical_2018} & 2D$^{\diamond}$ \cite{chang_2012} and 3D \cite{gerber_three-dimensional_2015, chang_magnetic_2016, bluschke_2018, kim_uniaxial_2018} & n/a  \\
		\hline 
		
		\hspace{0.05cm} Bi$_2$Sr$_2$CuO$_{6+\delta}$ (Bi-2201) & \hspace{0.05cm} 0.33 -- 0.29 \cite{Comin2014,peng_re-entrant_2018}  & \hspace{0.05cm} $\sim$0.3 \cite{Comin2014} & 2D \cite{peng_re-entrant_2018}  & Yes \cite{peng_re-entrant_2018}  \\
		\hline 

		\hspace{0.05cm} Bi$_2$Sr$_2$CaCu$_2$O$_{8+\delta}$ (Bi-2212) & \hspace{0.05cm} 0.33 -- 0.25 \cite{Eduardo2014_CDW_BISCCO}  & \hspace{0.05cm} $\sim$0.25 \cite{chaix_dispersive_2017} & 2D$^*$ & No \cite{YuHe_IXS_2018}  \\
		\hline 
		
		\hspace{0.05cm} ${\mathrm{HgBa}}_{2}{\mathrm{CuO}}_{4+\ensuremath{\delta}}$ (Hg-1201) & \hspace{0.05cm} 0.30 -- 0.265 \cite{Tabis2014,Tabis_PRB_2017}  & \hspace{0.05cm} $\in(\sim0.2,\sim0.3)$~\cite{Tabis2014} & 2D~\cite{Tabis_PRB_2017} & Unknown  \\
	    \hline 

		\hspace{0.05cm} Electron doped & \hspace{0.05cm} 0.16 -- 0.28 \cite{daSil15,da_silva_neto_coupling_2018,Jang2017}  & \hspace{0.05cm} 0.16 -- 0.28 \cite{daSil15,da_silva_neto_coupling_2018,Jang2017} & 2D$^*$ & Unknown  \\
		
		\hline \hline
	\end{tabular}
	\caption{\textbf{Summary of the key properties of charge order in cuprate families.} $^{\dagger}$: range of $q$-vectors denoted in order of increasing dopant concentration. $^{\diamond}$: phase-correlated along $c$-axis, with peaks at half-integer values of $L$, the out-of-plane reciprocal lattice units. $^*$: the 2-dimensionality of these systems is inferred but has not been directly confirmed.}
	\label{tab:CDW_summary}
\end{table*}

Bi-based cuprates were studied both by RXS and STS, and the ordering wavevectors inferred from both methods were found to be quantitatively consistent~\cite{Comin2014,Eduardo2014_CDW_BISCCO}. 
The length of the ordering wavevector decreases with increasing doping in all hole-doped systems except in the La$_2$CuO$_4$ system, and it increases with increasing electron doping, qualitatively consistent with the doping evolution of the Fermi surface diameter (Fig.~\ref{fig:q_vector}). A quantitative comparison between RXS and angle-resolved photoemission (ARPES) data on single layer Bi$_2$Sr$_2$CuO$_{6+\delta}$ showed that the length of the charge-ordering wavevector matches the distance between the tips of the ``Fermi arcs'' that are universally observed in the pseudogap regime of moderately hole-doped cuprates~\cite{Comin2014,Eduardo2014_CDW_BISCCO}. However, ARPES data do not indicate the formation of a gap at the Fermi-arc tips in the charge-ordered state~\cite{reber_origin_2012}, indicating that simple modifications of the classical CDW picture do not describe the data. Recent Raman spectroscopy experiments have revealed similar energy scales of the competing phases (charge order, superconductivity, and the pseudogap)~\cite{loret_intimate_2019}.

In the La$_2$CuO$_4$ system, the low-temperature charge-ordering wavevector increases with increasing doping, opposite to all other hole-doped cuprates. This behavior has been explained as a consequence of the proximity of independent spin and charge instabilities with nearly commensurate wave vectors \cite{nie_vestigial_2017}. A lock-in of the two wavevectors as a function of temperature observed by recent x-ray and neutron diffraction experiments supports this scenario~\cite{miao_RIXS_2017}. In other hole-doped systems including especially YBa$_2$Cu$_3$O$_{6+x}$, the wavevectors characterizing low-energy spin and charge fluctuations are quite different \cite{Blanco2013}, so that such a lock-in is less favorable.

The ionic displacement patterns associated with charge order have been investigated by both resonant and non-resonant x-ray diffraction. Most of the evidence reported so far indicates that the charge is concentrated in the center of the Cu-O-Cu bonds, with an anti-correlation of the amplitudes along the two orthogonal axes of the square lattice~\cite{Comin2015_NatMat,Comin2015_Science,Forgan2015}. This pattern, which is often referred to as ``$d$-wave'' charge order, is also consistent with STS observations \cite{Fujita2014,hamidian_2016} and with the results of calculations based on the 2D single-band Hubbard model ~\cite{Met2010, Sachdev2013}. However, different degrees of admixture of other symmetry components have also been reported \cite{Achkar2016,McMahon2019}, which is not surprising in view of the complex structure of real cuprate compounds and the significant displacement of ions outside the CuO$_2$ planes indicated by a full refinement of the x-ray diffraction pattern of underdoped YBa$_2$Cu$_3$O$_{6+x}$~\cite{Forgan2015}.

\subsection{Domains and influence of disorder.}

The low-temperature correlation lengths of the charge-ordered state vary greatly among the different cuprate families, ranging from a few unit cells in electron-doped and Bi- and Hg-based hole-doped superconductors all the way to several tens of lattice spacings in YBa$_2$Cu$_3$O$_{6+x}$ (Table~\ref{tab:corr_lengths}). This is generally consistent with the propensity for chemical disorder in these systems, which is minimal in YBa$_2$Cu$_3$O$_{6+x}$ because of its regular array of oxygen dopant ions. The shorter correlation lengths observed in other hole- and electron-doped cuprates are only weakly $T$-dependent, indicating that disorder limits the growth of charge-ordered domains. Short-range charge order with $T$-independent correlation length may be amenable to a description in terms of defect-induced ``Friedel oscillations'' akin to the ones that have been observed in weakly correlated metals~\cite{PhysRevB.93.205117}.

This description seems inadequate, however, for the charge-ordered states in YBa$_2$Cu$_3$O$_{6+x}$ and La$_2$CuO$_4$ whose correlation lengths increase strongly upon cooling, indicating an incipient critical divergence that is cut off at the onset of superconductivity. This behavior suggests the relevance of many-body interactions in driving charge ordering. At the same time, non-resonant inelastic x-ray scattering (IXS) with very high energy resolution indicates that the diffraction peaks corresponding to quasi-2D charge-order in YBa$_2$Cu$_3$O$_{6+x}$ have an energy width of at most $\sim 100$ $\mu$eV, which implies that the order is truly static, albeit short-range in space~\cite{le_tacon_inelastic_2014}. This finding is closely related to the ``central peak'' studied by Roger Cowley in soft-mode-driven structural phase transitions. Cowley proposed a model according to which lattice defects, which are unavoidable in real materials, pin some of the soft critical vibrations, resulting in an inhomogeneous state close to the transition~\cite{Cowley1996_STO,Cowley_2006}. A related situation arises from the confluence of disorder and interactions in YBa$_2$Cu$_3$O$_{6+x}$.  

Experiments sensitive to fluctuations with lower energies and larger time scales have confirmed the presence of static short-range charge order. In particular, nuclear magnetic resonance experiments revealed a line broadening attributable to charge ordering at a temperature comparable to the onset seen by x-rays~\cite{Wu2013,Wu2015}. Even longer time scales are probed by coherent x-ray scattering, which was recently deployed on La$_{2-x}$Ba$_{x}$CuO$_{4}$~\cite{chen_remarkable_2016}. These measurements confirmed the static nature of the charge order at all temperatures where the corresponding peak can be detected. Moreover, the same technique unveiled a memory effect of the charge order domains, shown in ~\cite{chen_charge_2019}. The authors observe how charge order domains forming at low temperature retain their internal structure even after cycling the temperature above the onset of the charge order (but below the high-temperature tetragonal-to-orthorhombic transition at 240\,K) and back down. However, when the sample is heated above 240\,K -- above which the system becomes tetragonal -- then cooled down, the charge order domains form in a different spatial configuration, and the memory is lost. The authors conclude that the structural features developing at the tetragonal-to-orthorhombic transition at 240\,K determine the charge order pinning landscape at low temperatures.

In principle, charge instabilities with $q$-vectors along the principal axes of the square lattice can either form ``single-$q$'' states with uniaxial order or a ``double-$q$'' state with checkerboard order. These two states are difficult to distinguish in scattering experiments on cuprates with tetragonal lattice structure, because the superposition of domains of two single-$q$ states resembles the diffraction pattern of the double-$q$ state. However, both RXS on YBa$_2$Cu$_3$O$_{6+x}$ (where the Cu-O chains generate an orthorhombic crystal structure) \cite{Comin2015_Science} and STS observations on Ca$_2$CuO$_2$Cl$_2$ and Bi-based superconductors~\cite{Fujita2014,hamidian_2016} indicate domains with uniaxial charge order~\cite{hoffman_four_2002,hanaguri_checkerboard_2004,kohsaka_intrinsic_2007,parker_fluctuating_2010}. The reconstructed Fermi surface inferred from quantum oscillation experiments, on the other hand, is more readily understood in models with biaxial charge order \cite{sebastian_QO_2012,chan_single_2016}. The energy balance between both states is subtle and may be influenced by disorder \cite{nie_vestigial_2017}. It is therefore conceivable that a checkerboard array forms under high magnetic fields, where the quantum oscillation experiments are carried out.

\begin{figure*}
  \centering
  \includegraphics[width=1\textwidth]
    {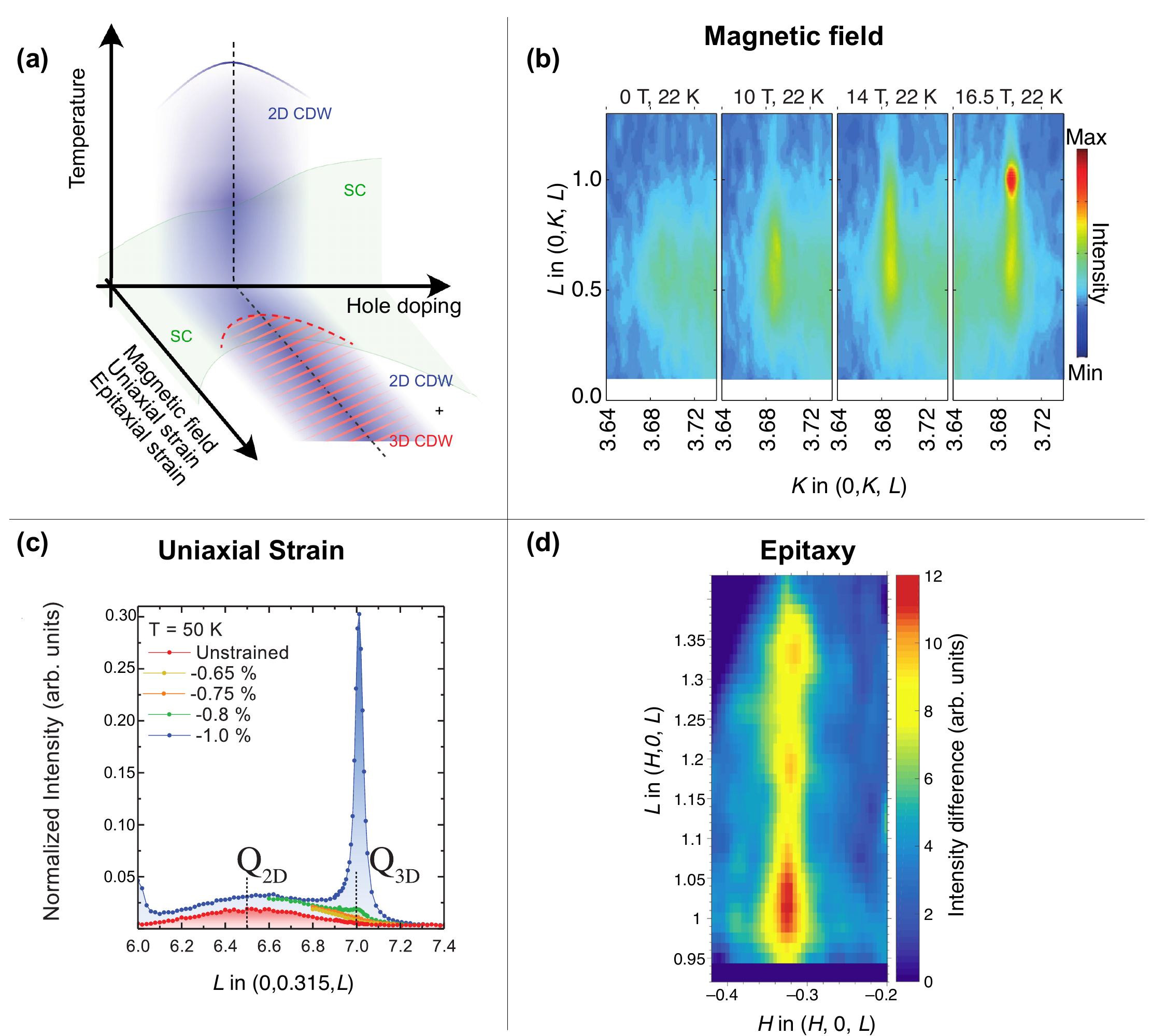}
     \caption{\textbf{Dimensionality of charge order in YBa$_2$Cu$_3$O$_{6+x}$.} (a) A generic phase diagram indicating the crossover from 2D to 3D charge order upon the application of magnetic field, uniaxial strain, and epitaxial strain (adapted from Ref.~\cite{Jang2016}). (b) With the application of a static~\cite{chang_magnetic_2016} and pulsed~\cite{gerber_three-dimensional_2015,Jang2016,Jang18} magnetic field of $\gtrsim$ 20\,T, the CO diffraction signal evolves from an elongated rod-like feature at 0 T along the out-of-plane reciprocal lattice direction $L$ (indicating a 2D phenomenon) to a sharp peak centered at integer values of $L$. The in-plane propagation vector (whose coordinates are given in square-lattice notation of YBa$_2$Cu$_3$O$_{6+x}$ and labeled $H,K$) remains unchanged (adapted from ~\cite{chang_magnetic_2016}). (c) $L$-scans of the CO peak revealing how the extended rod evolves into a sharp peak at $L=1$ upon the application of up to uniaxial strain (adapted from~\cite{kim_uniaxial_2018}). (d) A reciprocal space map of a YBa$_2$Cu$_3$O$_{6+x}$ film measured by RXS indicating both a rod feature and a sharp peak at $L=1$ (adapted from~\cite{Bluschke2019}).} \label{fig:Dimensionality}
\end{figure*}

An extreme case of interaction-driven charge order is the three-dimensional (3D) charge-ordered state, which was recently discovered by x-ray experiments on YBa$_2$Cu$_3$O$_{6+x}$ in high magnetic fields~\cite{gerber_three-dimensional_2015,Jang2016,Jang18,chang_magnetic_2016} and under high uniaxial strain~\cite{kim_uniaxial_2018}. This state is heralded by sharp Bragg peaks centered at integer values of the momentum transfer perpendicular to the CuO$_2$ layers, in contrast to the diffraction rods indicating quasi-2D charge order under ambient conditions (Fig.~\ref{fig:Dimensionality}). Relatedly, resonant soft x-ray scattering has recently demonstrated a 3D charge order peak in YBa$_2$Cu$_3$O$_{6+\delta}$ films grown epitaxially on SrTiO$_3$ in the absence of magnetic fields  ~\cite{bluschke_2018} (Fig.~\ref{fig:Dimensionality}(d)).
This state is genuinely long-range ordered, with domain sizes of $\sim 100$ lattice spacings in-plane and $\sim 10$ spacings out-of-plane inferred from the width of the Bragg reflections (Table~\ref{tab:corr_lengths}). The field- and strain-induced 3D charge order is uniaxial, with a propagation vector that is directed along the orthorhombic axis parallel to the Cu-O chains. The length of the ordering wavevector matches the one of the 2D charge ordered state in the same materials in the absence of strain and magnetic fields.  

The universality of 3D charge order and its role in enabling the exceptionally pronounced quantum oscillations in underdoped YBa$_2$Cu$_3$O$_{6+x}$ are subjects of current investigation. This state may be favored by the electronically active Cu-O chains in the YBa$_2$Cu$_3$O$_{6+x}$ crystal structure, which generate an additional open sheet of the Fermi surface that is not present in most other cuprates~\cite{bluschke_2018}. The complex electronic structure generated by the CuO chains, its modification by uniaxial or epitaxial strain, and its influence on charge ordering phenomena are interesting subjects of future theoretical and experimental research.

The data described above strongly suggests that disorder is central to the behavior of the charge order.
It indicates possible realizations of both the 2D and 3D random field XY models \cite{Imry1975,Aharony1983}. Note that for the XY model, random fields are much more destructive of the order in 2D compared with 3D. In both cases it prevents the achievement of true long-range order. This difference in the effects of structural (i.e. random field) disorder in two and three dimensions could explain why the quasi-2D charge order is of much shorter range compared to that in the 3D correlated charge ordered state.

\subsection{Charge order in the phase diagram.}

Table~\ref{tab:CDW_summary} and Fig.~\ref{fig:phase_diagram} summarize the charge ordering phenomenology across various cuprate systems. Most research on charge order has thus far focused on the ``pseudogap'' regime, which sets in below the temperature $T^*$ marked in the phase diagram of Fig.~\ref{fig:phase_diagram}. Two major explanations have been advanced for the origin of the pseudogap: (i) a crossover due to electronic correlations that also generate antiferromagnetic short-range order, and (ii) a phase transition with a broken symmetry that is ``hidden'' to most experimental probes. The former explanation holds for both hole-doped and electron-doped superconductors, where the onset of the pseudogap is less well defined \cite{fujimori}. Comprehensive x-ray scattering surveys of YBa$_2$Cu$_3$O$_{6+x}$ and La$_{2-x}$Sr$_{x}$CuO$_{4}$, whose charge-ordered states are driven by many-body interactions, indicate a gradual onset of diffuse features associated with charge ordering. Although there is some ambiguity in the definition of the corresponding onset temperature, the x-ray data clearly indicate that $T_{CO}$ is lower than $T^*$ and displays a different doping dependence. Whereas $T^*$ decreases monotonically with increasing hole doping, $T_{CO}$ subtends a ``dome'' in the phase diagram analogous to the superconducting dome, but centered at a lower doping level~\cite{Blanco2013,Hucker2014}. This finding clearly rules out models according to which static charge order is the origin of the pseudogap.

Based on this phenomenology, charge order has been widely viewed as an instability secondary to the one driving the formation of the pseudogap. In particular, if the formation of the pseudogap and Fermi arcs can be understood as a consequence of strong correlations, models of charge order can take these features of the electronic structure as a starting point. However, recent experiments have challenged this viewpoint. The first challenge came from a resonant x-ray scattering study of overdoped (Bi,Pb)$_{2.12}$Sr$_{1.88}$CuO$_{6+\delta}$, which uncovered diffraction features remarkably similar to the ones in the underdoped regime of the same material, and with wavevectors consistent with a simple extrapolation of the doping dependent trend established there~\cite{peng_re-entrant_2018}. The maximum amplitude of these features is observed near the Lifshitz point, where the Fermi surface passes through a van Hove singularity and its topology changes from hole-like to electron-like. The charge-order correlation length inferred from their momentum-space width decreases as a function of increasing temperature, akin to findings in the underdoped regime, but the onset temperature is well above room temperature. Since the effective strength of the electronic correlations decreases in highly overdoped cuprates, a different mechanism has to be invoked to explain the exceptionally robust charge-ordered state in overdoped Bi$_2$Sr$_2$CuO$_{8+\delta}$. A particular scenario is based on the high density of states at the Fermi level generated by the van Hove singularity, which may be conducive to charge density wave formation -- in analogy to the density-of-states enhancement originating from the narrowing of the valence band by many-body interactions in the underdoped regime. Future work should aim for a consistent description of the fermiology and the structure of the charge-ordered state in the overdoped regime~\cite{peng_re-entrant_2018}. In particular, charge ordering is expected to induce a reconstruction of the Fermi surface that should be observable by ARPES. It is also important to assess the universality of this state in the cuprates. Initial experiments on highly overdoped Bi$_2$Sr$_2$CaCu$_2$O$_{8+\delta}$ did not uncover evidence of charge ordering \cite{YuHe_IXS_2018}.

A second challenge for the prevailing view of charge ordering as a secondary instability came from a recent report by Arpaia \textit{et al.}~\cite{arpaia_dynamical_2018} who attributed the nearly temperature-independent, sloping background detected by various resonant scattering experiments on YBa$_2$Cu$_3$O$_{6+x}$ ~\cite{Blanco2013, Blanco2014, Comin2014, Eduardo2014_CDW_BISCCO, Frano2016} to a charge fluctuation signal covering a broad region of reciprocal space (Figure~\ref{fig:phase_diagram}). The wavevector characterizing this signal is close to, but distinct from the diffraction features heralding an incipient long-range charge-ordered state discussed above. Other distinct characteristics are the nonzero energy width of the broad feature, which indicates fluctuating rather than static order, and its persistence up to at least room temperature. If this finding can be generalized to other families of cuprates, it may indicate the presence of dynamical nearest-neighbor bond correlations over a temperature range comparable to that of the antiferromagnetic spin correlations, and possibly reflecting the same tendency to form singlets.

\subsection{Competition and coexistence with superconductivity.}

A priori one might expect a strong competition between charge order directed along the Cu-O-Cu bonds and $d$-wave superconductivity which also exhibits maximal strength in this direction. This expectation is indeed confirmed by several distinct experimental signatures. First, the amplitude and correlation length inferred from the charge-ordering reflections in YBa$_2$Cu$_3$O$_{6+x}$ and La$_{2-x}$Sr$_{x}$CuO$_{4}$ exhibit cusp-like maxima at $T_c$, indicating that the charge order is weakened by the onset of superconductivity. Second, x-ray scattering experiments under high magnetic fields showed that the reflections sharpen and their amplitude is restored, presumably because superconductivity is weakened by orbital depairing~\cite{Blanco2013,chang_2012,chang_2012}. Third, the superconducting dome in the phase diagram exhibits a dip at doping levels $p \sim 0.12$ where the amplitude and correlation length of charge order is maximal~\cite{Blanco2014,Hucker2014}. This effect is most pronounced in La$_{2-x}$Ba$_{x}$CuO$_{4}$ where $T_c$ is suppressed almost entirely by the striped phase with coexisting charge and spin order \cite{Hucker_LBCO_PRB_2011}. Finally, the enhancement of superconductivity via ultrafast phonon pumping~\cite{fausti_light-induced_2011,Kaiser_light-induced_2014,Nicoletti2014} has been connected to a suppression of the charge order peak intensity in both La$_{2-x}$Ba$_{x}$CuO$_{4}$~\cite{Forst_LBCO_CO_melting} and in YBa$_2$Cu$_3$O$_{6+x}$~\cite{Forst_YBCO_CO_melting}.

\begin{center}
\begin{table*}[htb]
	\begin{tabular}{ || p{3cm} |  p{4cm} | p{3cm} | p{4cm}  ||  }
		\hline \hline
		\hspace{0.05cm}\textbf{State} & \hspace{0.05cm} \textbf{In-plane $\xi$ (\r{A})} & \hspace{0.05cm} \textbf{Out-of-plane $\xi$ at $L=\frac{n}{2}$ (\r{A})}   &   \textbf{Out-of-plane $\xi$ at $L=n$ (\r{A})} \hspace{0.15cm}  \\

		\hline \hline
		\hspace{0.05cm} Unperturbed & \hspace{0.05cm} 60-70 \cite{Blanco2014,Hucker2014}; 95 \cite{chang_2012}; 50 \cite{Hucker2014}  & \hspace{0.05cm} ~10 \cite{chang_2012} & n/a  \\
		\hline
		\hspace{0.05cm} Magnetic field & \hspace{0.05cm} 100 at $H < 20$\,T, 180 at $H > 28$\,T \cite{gerber_three-dimensional_2015,Jang2016} at $L=1$; 100 at $H = 16$\,T at $L=\frac{n}{2}$ \cite{chang_magnetic_2016}; 300-400 at $H =16$\,T at $L=1$ \cite{chang_magnetic_2016} &  \hspace{0.05cm} $\sim$10 \cite{chang_magnetic_2016} & 34 at $H = 20$\,T, 50 at $H = 28$\,T \cite{gerber_three-dimensional_2015, Jang2016}; 47 at $H = 16$\,T  \cite{chang_magnetic_2016} \\
		\hline 
		\hspace{0.05cm} Uniaxial strain & \hspace{0.05cm} 310 at 1\% \cite{kim_uniaxial_2018} & \hspace{0.05cm} $\sim$10 \cite{kim_uniaxial_2018} & 94 at 1\% \cite{kim_uniaxial_2018}\\
		\hline
		\hspace{0.05cm} Epitaxial strain & \hspace{0.05cm} 80-100 \cite{bluschke_2018} & \hspace{0.05cm} $\sim$10 \cite{bluschke_2018} & 50-60 \cite{bluschke_2018} \\
		\hline \hline
	\end{tabular}
	\caption{\textbf{Correlation lengths of 2D and 3D charge orders in YBa$_2$Cu$_3$O$_{6+x}$.} A comparison of the correlation lengths $\xi$ of the charge order in YBa$_2$Cu$_3$O$_{6+x}$ under different experimental conditions. }
	\label{tab:corr_lengths}
\end{table*}
\end{center}

The strength of these signatures varies greatly among different materials, reflecting the degree of lattice disorder that pins the charge order correlations. In Bi- and Hg-based superconductors, the superconductivity-induced anomalies of the charge-ordering reflections are minimal \cite{Eduardo2014_CDW_BISCCO,Tabis2014}. Recent experiments on La$_{2}$CuO$_{4}$ systems did not show an enhancement of the charge order reflections in magnetic fields~\cite{Hucker2013,Zwiebler2016,BlancoCanosa2018}, likely due to pinning by lattice distortions. Even in YBa$_2$Cu$_3$O$_{6+x}$, the reflections indicative of quasi-2D short-range order under ambient conditions are not entirely suppressed below $T_c$, indicating inhomogeneous coexistence of superconductivity and charge order. It is interesting to note the analogy between lattice defects, which pin incommensurate charge-order fluctuations, and spinless Zn impurities, which are known to pin low-energy incommensurate spin fluctuations in the cuprates \cite{suchaneck_incommensurate_2010}. RXS experiments on Zn-substituted YBa$_2$Cu$_3$O$_{6+x}$ revealed that these impurities also suppress charge order~\cite{Blanco2013}, evidencing a three-phase competition between superconductivity, incommensurate magnetic order, and incommensurate charge order in this system.

The only instance in which the competition between charge order and superconductivity seems complete, without inhomogeneous coexistence from pinning to lattice defects, is the 3D charge-ordered state generated in YBa$_2$Cu$_3$O$_{6+x}$ under uniaxial strain \cite{kim_uniaxial_2018}. Non-resonant x-ray experiments showed that the reflections from this state are completely obliterated in the superconducting state, underscoring its nature as a genuine thermodynamic phase. This observation points out new perspectives to study the interplay between interaction-driven quantum phases with minimal influence of disorder.

The near-degeneracy of the charge-ordered and superconducting phases in underdoped cuprates has stimulated theories that describe the fluctuations in the pseudogap states in terms of a composite order parameter that comprises both superconducting and charge-order components~\cite{Pepin2014,hayward_angular_2014}. While these models reproduce both the gradual onset of the quasi-2D charge-order reflections and the cusp of their intensity at $T_c$, deviations from the corresponding predictions are apparent both at low and at high temperatures, presumably due to the influence of disorder and/or electron phonon interactions. We also note predictions of a charge modulation with a wavevector twice as long as the charge-ordering vector from pair-density-wave correlations~\cite{Lee_PDW}, which have received some support from recent scanning tunneling microscopy experiments~\cite{Edkins2019}. Observations of such reflections in x-ray experiments have thus far not been reported.

\subsection{Collective charge dynamics and phonon anomalies.}

As in the case of spin excitations, measurements of the dynamical structure factor associated with collective charge fluctuations have the potential to uncover detailed information about the interactions that drive the charge ordering instability. Such excitations have recently been directly observed via resonant inelastic x-ray scattering (RIXS) at energies up to 60 meV, and were shown to hybridize strongly with phonons via electron-phonon interactions so that they are difficult to separate from the phonon spectrum~\cite{chaix_dispersive_2017}. Magnetic RIXS experiments on electron-doped Nd$_{2-x}$Ce$_{x}$CuO$_{4}$ have uncovered anomalies in the paramagnon modes at the wavevector characterizing charge order, indicating a dynamical coupling between collective spin and charge excitations at energies up to about 300 meV~\cite{daSil16,da_silva_neto_coupling_2018}. This observation is consistent with the high temperature scale of the broad charge fluctuation signal from quasielastic RIXS experiments discussed above. 


\begin{figure*}
  \centering
  \includegraphics[width=0.95\textwidth]
    {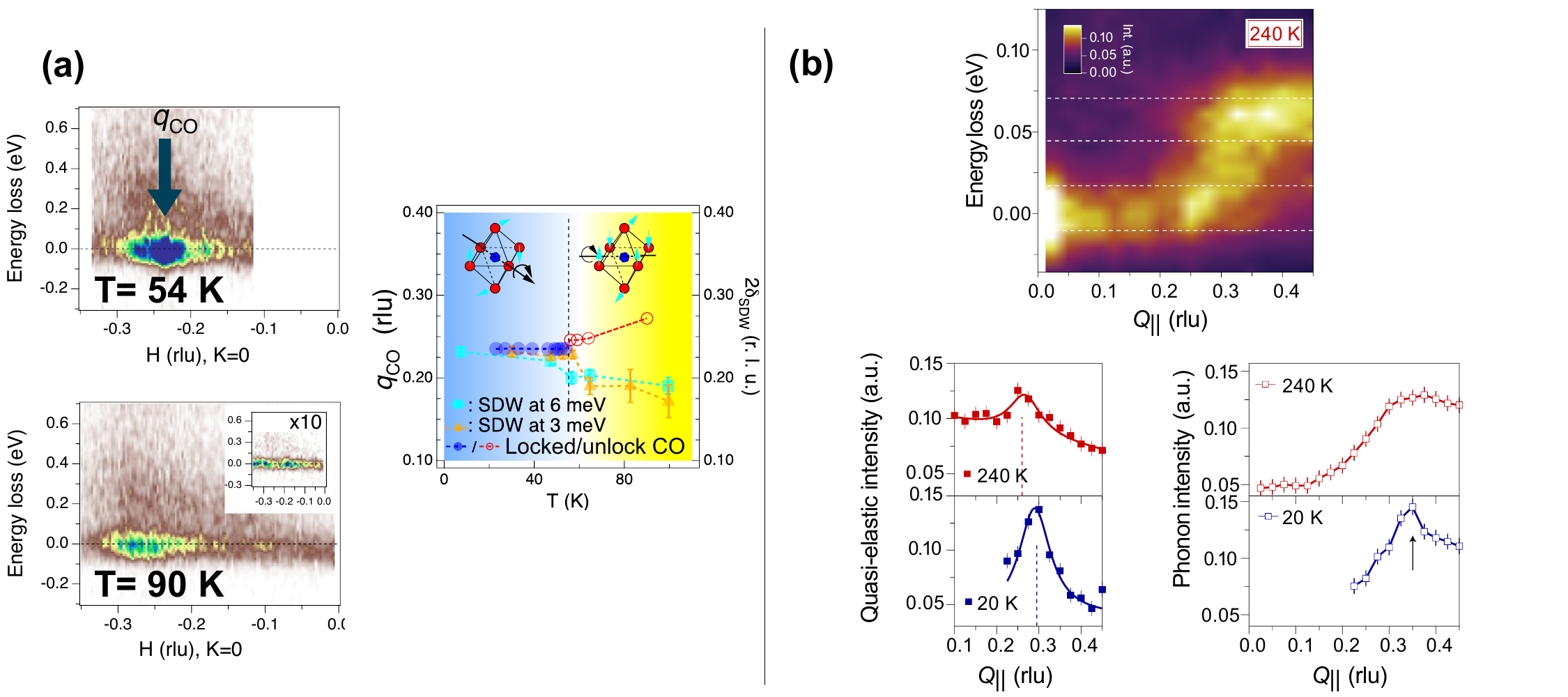}
    \caption{\textbf{RIXS studies of trends in charge order $q$-vector.} (a) RIXS measurements from the La$_2$CuO$_4$ based system reveal a shift in the in-plane $q$-vector (with coordinates $H,K$ in square-lattice notation) of the low-temperature CO peak at zero energy loss (top left) with respect to that of high-temperature fluctuations (bottom left). The right panel shows the decoupling of the incommensurability of the CO and spin density wave in the high-temperature phase (yellow region). Upon warming, the shift is from $\sim$0.25\,rlu to $\sim$0.3\,rlu. The black dashed line at 54 K corresponds to the structural phase transition~\cite{Fink_LESCO_1} (adapted from~\cite{miao_RIXS_2017}). (b) The reverse trend is seen by RIXS in Bi$_2$Sr$_2$CaCu$_2$O$_{8+\delta}$. The top panel shows a RIXS intensity map taken at 240\,K revealing a broad feature at 60\,meV around $\sim$0.35\,rlu. The dotted white lines denote the energy windows used for integration as follows. In the bottom left panel, the quasi-elastic intensity (integrating the energy range near zero energy loss) is plotted at two temperatures. The dashed lines indicate the $q$-vector of the CO. A shift from $\sim$0.3\,rlu to $\sim$0.25\,rlu upon warming is observed. In the bottom right panel, the inelastic intensity (integrating over an energy window that spans the 60\,meV feature) is plotted versus the in-plane momentum transfer $Q_{||}$ along the Cu-O bond direction (adapted from~\cite{chaix_dispersive_2017}). } 
    \label{fig:RIXS}
\end{figure*}

Recent RIXS experiments on La$_{2-x}$Ba$_{x}$CuO$_{4}$ ~\cite{miao_RIXS_2017,Miao2019} have been conducted with the aim of obtaining a unified picture of the doping-dependent $q$-vector across families (Fig.~\ref{fig:RIXS}). The $q$-dependent RIXS signal as a function of temperature (Figure~\ref{fig:RIXS}(a)) shows a shift from the commensurate value of $\sim$0.25\,reciprocal lattice units (rlu) at low temperatures in the static ordered state to 0.27\,rlu above 55\,K. As the $q$-vector value approaches that of the other families, charge and spin correlations decouple above 55\,K. Since the scattered intensity vanishes rapidly above 80\,K, it is not clear whether the upwards trend in the $q$-vector reaches 0.3\,rlu at higher temperatures. 

However, the trend of $q$-vector versus temperature of the charge fluctuations is also material-dependent (Fig.~\ref{fig:RIXS}). A recent RIXS report on double-layer Bi$_2$Sr$_2$CaCu$_2$O$_{8+\delta}$ showed an opposite trend in the $q$-vector as function of temperature~\cite{chaix_dispersive_2017}. At low temperatures, the $q$-vector of the elastic peak is roughly 0.3\,rlu but shifts to lower values upon heating, approaching ~0.25\,rlu. Detailed theoretical work is required to assess the origin of the temperature dependence of the characteristic $q$-vector for the charge response and its relationship to the electronic structure and lattice dynamics of specific compounds.
 
\begin{figure*}
  \centering
  \includegraphics[width=0.95\textwidth]
    {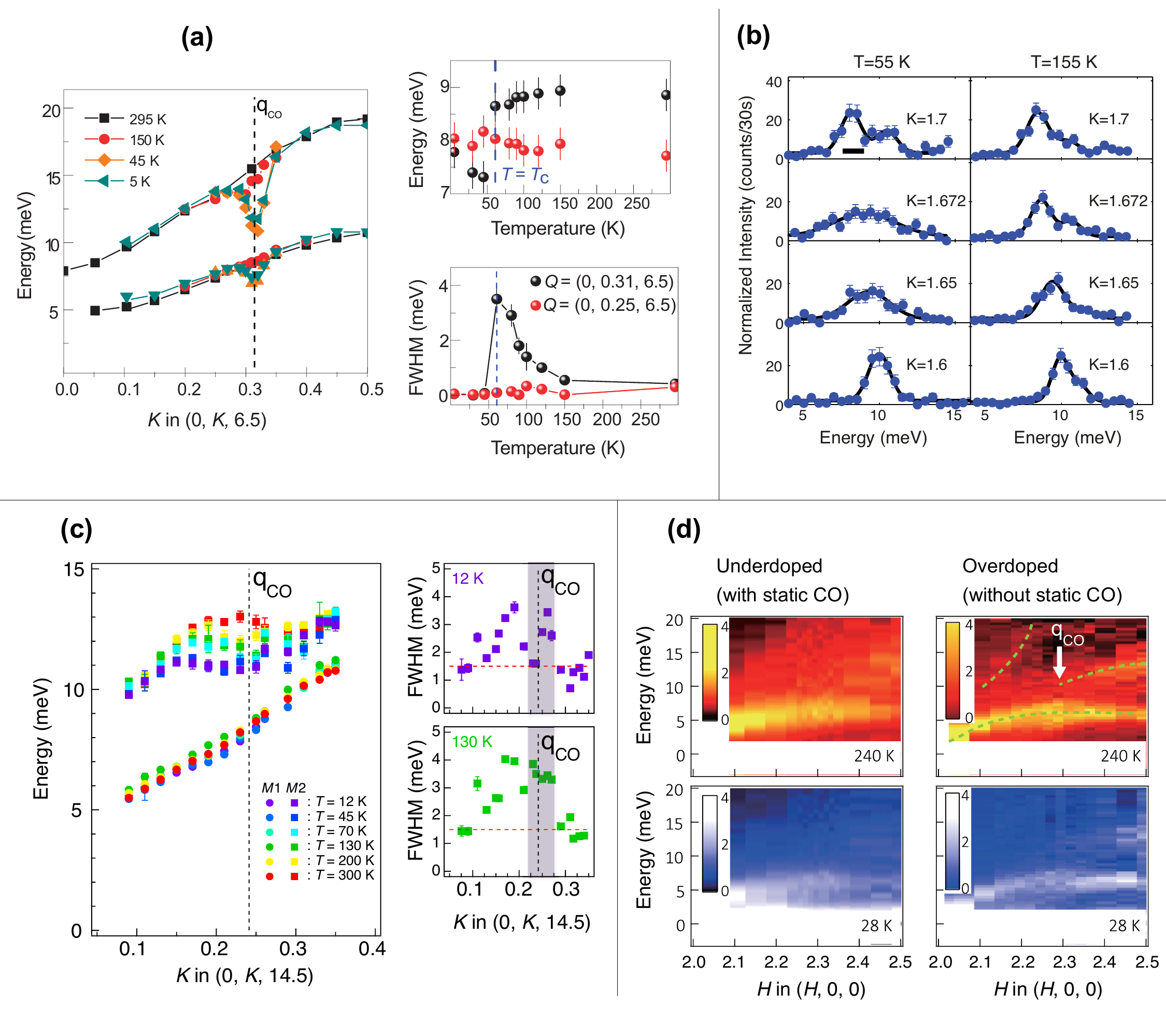}
    \caption{\textbf{Phonon anomalies revealed by IXS.} (a) In YBa$_2$Cu$_3$O$_{6+x}$, two low-energy phonon branches reveal dips in their dispersion curves at $\sim$0.3\,rlu near the CO $q$-vector (denoted by dotted black line) at low temperatures (left panel). The phonons soften (top right) and broaden (bottom right) at the onset of superconductivity (adapted from~\cite{le_tacon_inelastic_2014}). (b) Another group observed the same broadening but not the softening (adapted from~\cite{blackburn_inelastic_2013}). (c) In La$_{2-x}$Ba$_{x}$CuO$_{4}$, similar phonon branches soften with decreasing temperature at $\sim$0.25\,rlu near the CO $q$-vector (denoted by dotted black lines). In addition, the full width at half maximum (FWHM) of the higher energy phonon broadens in the region of the charge order $q$-vector at both low (top right) and high (bottom right) temperatures, potentially as a precursor of the CO (adapted from~\cite{miao_IXS_2018}). (d) A similar broadening was observed in Bi$_2$Sr$_2$CaCu$_2$O$_{8+\delta}$ at all temperatures and for underdoped (with static CO) and overdoped (without static CO) samples. Green dotted lines are guide to the eye. The white arrow marker indicates the CO $q$-vector (adapted from~\cite{YuHe_IXS_2018}). The momentum coordinates in all panels are given in square-lattice notation and labeled $(H, K, L)$.
    }\label{fig:IXS_summary}
\end{figure*}

Imprints of collective charge fluctuations on the phonon spectrum have also been observed by inelastic neutron scattering and high-resolution non-resonant inelastic x-ray scattering, which are sensitive to the ionic positions via interactions with nuclei and core electrons, respectively (Figure~\ref{fig:IXS_summary}). Neutron scattering experiments on La$_{2-x}$Sr$_{x}$CuO$_{4}$ and YBa$_2$Cu$_3$O$_{6+x}$ found pronounced anomalies at the charge ordering wave vector of the Cu-O-Cu bond-stretching and bond-bending phonons that are common to all cuprates \cite{reznik_photoemission_2008,Raichle2011,Park_collective_mode_2014}. High-resolution inelastic x-ray scattering experiments have detected sharp anomalies also in acoustic and low-energy optical phonon branches of these materials. In underdoped YBa$_2$Cu$_3$O$_{6+x}$, these anomalies involve pronounced phonon broadening (up to $\sim 40\%$ of the phonon energy) and sharp Kohn anomalies in the dispersion above and below $T_c$, respectively~\cite{le_tacon_inelastic_2014, blackburn_inelastic_2013,Souliou2018}. The complete softening of a phonon was seen at the 3D charge ordering transition of YBa$_2$Cu$_3$O$_{6+x}$ under uniaxial strain \cite{kim_uniaxial_2018}.

The connection between the lattice dynamics and the charge order has also been explored by means of IXS in cuprate families beyond YBa$_2$Cu$_3$O$_{6+x}$ (Figure~\ref{fig:IXS_summary}(c)). In the La$_2$CuO$_{4}$ family, a broadening of an acoustic phonon branch at 10-12\,meV was recently observed~\cite{miao_IXS_2018}. The range of $q$-values over which the broadening occurs is temperature dependent. At low temperatures the broadening is centered narrowly around $q_{CO}\sim 0.23$\,rlu. The linewidth of the phonon mode decreases when cooling through the charge order onset temperature. However, at 300\,K the reciprocal space region where the broadening occurs spans a much wider range between [0.2-0.3]\,rlu. This result, consistent with the RIXS report we discussed previously~\cite{miao_RIXS_2017,Miao2019}, indicates a commonality between the La$_2$CuO$_{4}$ system and the other cuprate families. 

Recent experiments on Bi$_2$Sr$_2$CaCu$_2$O$_{8+\delta}$ found phonon anomalies at the charge-ordering wavevector that persist above the onset of static charge-order, possibly signalling a soft-phonon precursor \cite{YuHe_IXS_2018}. However, the temperature-independent broadening was also observed in the overdoped regime, far away from the charge order as is currently established, and alternative explanations based on conventional lattice dynamics have also been proposed~\cite{Merritt2019}. Another study using inelastic neutron scattering confirmed the broadening in the single-layer compound Bi$_2$Sr$_2$CuO$_{8+\delta}$ at optimal doping~\cite{bonnoit_probing_2012} (where static charge order has also not been reported). The nature of this phonon anomaly is still an open question which could be resolved by modern RIXS experiments with higher spectral resolution that are becoming capable of directly determining the electron-phonon coupling~\cite{rossi_experimental_2019}. 

Taken together, these experiments imply a significant influence of electron-phonon interactions on the energetics of charge ordering, in addition to the electronic energies that have been considered in most of the theoretical work on this issue. Some of the materials-specific variations of charge order in the cuprates may well reflect variations of the electron-phonon coupling in different lattice structures. We note that the recent observations on classical CDW materials have also been interpreted as evidence of an important role of momentum-dependent electron-phonon interactions on the formation of the charge density wave~\cite{Weber}.


\subsection{Outlook}

Over the past few years, a multitude of spectroscopic methods including resonant x-ray scattering, high-resolution non-resonant inelastic x-ray scattering, and scanning tunneling spectroscopy has yielded a new ``universal'' phase diagram of the cuprates that now includes a prominent charge-ordering dome centered at moderate doping levels. 
Recent experiments have begun to add multiple new dimensions to this diagram through tunable control parameters such as high magnetic fields, uniaxial and hydrostatic pressure, and intense THz light fields. These experiments are still at a stage of rapid development and will likely yield fresh insight in the near future. In particular, {\it in situ} combinations of spectroscopic, transport, and thermodynamic measurements 
under the influence of external parameters have the potential to elucidate the influence of microscopic charge correlations on the macroscopic properties in a much more systematic fashion than prior work monitoring narrow two-dimensional segments of this multidimensional phase diagram.

Another development that has only just begun takes advantage of recent progress in the synthesis of thin films and heterostructures to systematically manipulate the electron system in the cuprates. In particular, the high sensitivity of RXS allows measurements on films that are only a few unit cells thick and thus adopt the lattice structure of the underlying substrate. For instance, RXS experiments on thin YBa$_2$Cu$_3$O$_{6+x}$ films with a tetragonal structure imposed by the SrTiO$_3$ 
substrate uncovered a robust 3D charge-ordered state that is not present in bulk YBa$_2$Cu$_3$O$_{6+x}$ under ambient conditions~\cite{bluschke_2018} (Fig.~\ref{fig:Dimensionality}(d)). In contrast to the state realized under high uniaxial strain, however, the intensity of the Bragg reflections characteristic of 3D charge order depends monotonically on temperature, without evidence of competition with superconductivity. This indicates a profound influence of the substrate on the electronic ground state of the cuprate film. Experiments on cuprate-manganate superlattices revealed a massive rearrangement of the 
orbital occupation and electronic structure at the hetero-interfaces \cite{chakhalian_orbital_2007} and an unexpectedly long-range influence of the interfaces on the electron-phonon interaction~\cite{driza_2012} and charge-ordering phenomena~\cite{Frano2016,He16}. Very recent experiments on related systems have led to the discovery of a highly unusual 
magnetic-field-induced insulator-superconductor transition, which may reflect an imprint of the field-sensitive charge-ordered state in the manganate onto the CuO$_2$ layers~\cite{Khmaladze2019}. These results indicate the vast potential of metal-oxide heterostructures as a platform for systematic manipulations of the electronic properties of 
the cuprates.

To fully realize this potential, these experimental developments must go hand in hand with advances in theoretical research, which have recently led to increasingly realistic descriptions of correlation effects in complex solids. Since the combination of disorder and correlations continues to present a formidable challenge, recent results on bulk 
crystals and superlattices with minimal disorder and full translational periodicity present new opportunities for ab-initio modeling. In particular, the discovery of 3D charge order in oxygen-ordered YBa$_2$Cu$_3$O$_{6+x}$ has established a model system in which both the host lattice and the electronic superstructure exhibit complete long-range order. This situation should be highly favorable for realistic modeling of the electron-phonon interaction and its influence on the stability of the 
charge-ordered state.

The ultimate aim of research on the cuprates is a microscopic theory of high-temperature superconductivity. In this context, the results collected over the past few years unambiguously demonstrate that static charge order of any kind (with and without coexisting spin order) competes against superconductivity. Important open questions are the role of soft collective charge fluctuations in proximity to the charge-ordered state, and the relevance of long-range entanglement generically associated with quantum criticality~\cite{Wang_Chubukov2015, Lederer2017}. Transport experiments on YBa$_2$Cu$_3$O$_{6+x}$ in high magnetic fields indicated two separate superconducting domes, which are associated with a divergence of the electronic mass \cite{Ramshaw2015}. The observation that these domes are centered near the end points of the stability range of quasi-2D charge order in YBa$_2$Cu$_3$O$_{6+x}$~\cite{Blanco2014,Hucker2014} suggests a possible supporting role of quantum-critical charge fluctuations for superconductivity, on top of the spin fluctuations that are known to favor $d$-wave 
superconductivity. However, a causal relationship could not be firmly established due to the influence of disorder, which leads to a gradual onset of quasi-2D charge order as a function of both temperature and doping and to inhomogeneous coexistence with superconductivity at low temperatures. Moreover, scattering methods probing the microscopic charge order and dynamics could thus far not be performed under the high-field conditions required for quantum transport experiments.

Finally, we note that a wide variety of compounds with less correlated electron systems also exhibit phase diagrams with proximate superconducting and charge-ordered phases, ranging from organic charge-transfer salts~\cite{clay_2019} all the way to structural homologues of the iron-pnictide high-temperature superconductors~\cite{allred_double_q_2016,Wasser2015,Lee_CDW_pnictides_2019,MingYi2018}. In each of these cases, the influence of collective charge fluctuations on the mechanism of superconductivity and their interplay with spin correlations and lattice vibrations remains unresolved. The recent advances in materials synthesis and experimental methodology we have discussed provide exciting perspectives for decisive progress on this central issue.


\subsection{Acknowledgements}

First of all, we would like to thank and honor Roger A. Cowley for his five decades of inspiring experimental and theoretical research in solid state physics.  There is almost no subject in our field where Roger has not written one of the foundational papers, be it on structural phase transitions, lattice dynamics, magnetic excitons, low dimensional, quantum magnetism, disordered systems, superconductivity and superfluidity. His papers are models of clarity and precision.  Roger was an ideal collaborator, modest and self-deprecating but at the same time both creative and brilliant. His passing was a great loss to our field but his legacy will last.

Work at Lawrence Berkeley National Laboratory was funded by the U.S. Department of Energy, Office of Science, Office of Basic Energy Sciences, Materials Sciences and Engineering Division under Contract No. DE-AC02-05-CH11231 within the Quantum Materials Program (KC2202). Work at DIPC is supported by IKERBASQUE and the MINECO of Spain through the project PGC2018-101334-A-C22.

We would like to thank Peter Abbamonte, Elizabeth Blackburn, Martin Bluschke, Lucio Braicovich, Johan Chang, Xiang Chen, Riccardo Comin, Andrea Damascelli, Mark Dean, Tom Devereaux, Giacomo Ghiringelli, Martin Greven, David Hawthorn, Stephen Hayden, Feizhou He, Yu He, Vladimir Hinkov, Markus H\"ucker, Steve Kivelson, Wei-Sheng Lee, Yuan Li, Toshinao Loew, Claudio Mazzoli, Matteo Minola, Marco Moretti, Eduardo da Silva Neto, Juan Porras, Christian Sch\"u{\ss}ler-Langeheine, Wojciech Tabis, Matthieu Le Tacon, John Tranquada, George Sawatzky, Enrico Schierle, Suchitra Sebastian, Padraic Shafer, Z.X. Shen, Ronny Sutarto, Eugen Weschke, and Ming Yi for fruitful discussions and collaborations.

\bibliographystyle{unsrt}

\providecommand{\latin}[1]{#1}
\makeatletter
\providecommand{\doi}
  {\begingroup\let\do\@makeother\dospecials
  \catcode`\{=1 \catcode`\}=2 \doi@aux}
\providecommand{\doi@aux}[1]{\endgroup\texttt{#1}}
\makeatother
\providecommand*\mcitethebibliography{\thebibliography}
\csname @ifundefined\endcsname{endmcitethebibliography}
  {\let\endmcitethebibliography\endthebibliography}{}

\end{document}